\documentclass[sigconf]{acmart}
\usepackage{graphicx} 

\usepackage{xcolor}
\newcommand{\modif}[1]{\textcolor{black}{#1}}

\title{How Fair is Software Fairness Testing?}

\author{Ann Barcomb}
\authornote{All authors contributed equally to this paper.}
\affiliation{
  \institution{University of Calgary}
  \city{Calgary}
  \country{Canada}
}
\email{ann.barcomb@ucalgary.ca}

\author{Mariana Pinheiro Bento}
\authornotemark[1]
\affiliation{
  \institution{University of Calgary}
  \city{Calgary}
  \country{Canada}
}
\email{mariana.pinheirobent@ucalgary.ca}

\author{Giuseppe Destefanis}
\authornotemark[1]
\affiliation{
  \institution{University College London}
  \city{London}
  \country{United Kingdom}
}
\email{g.destefanis@ucl.ac.uk}

\author{Sherlock Licorish}
\authornotemark[1]
\affiliation{
  \institution{University of Otago}
  \city{Dunedin}
  \country{New Zealand}
}
\email{sherlock.licorish@otago.ac.nz}

\author{Cleyton Magalhães}
\authornotemark[1]
\affiliation{
  \institution{Federal Rural University of Pernambuco}
  \city{Recife}
  \country{Brazil}
}
\email{cleyton.vanut@ufrpe.br}

\author{Ronnie de Souza Santos}
\authornotemark[1]
\affiliation{
  \institution{University of Calgary}
  \city{Calgary}
  \country{Canada}
}
\email{ronnie.desouzasantos@ucalgary.ca}

\author{Mairieli Wessel}
\authornotemark[1]
\affiliation{
  \institution{Radboud University}
  \city{Nijmegen}
  \country{Netherlands}
}
\email{mairieli.wessel@ru.nl}

\begin{abstract}
Software fairness testing is a central method for evaluating AI systems, yet the meaning of fairness is often treated as fixed and universally applicable. This vision paper positions fairness testing as culturally situated and examines the problem across three dimensions. First, fairness metrics encode particular cultural values while marginalizing others. Second, test datasets are predominantly designed from Western contexts, excluding knowledge systems grounded in oral traditions, Indigenous languages, and non-digital communities. Third, fairness testing raises ethical concerns, including the reliance on low-paid data labeling in the Global South, and associated with this, the environmental costs of training and deploying large-scale models, which disproportionately affect climate-vulnerable populations. Addressing these issues requires rethinking fairness testing beyond universal metrics and moving toward evaluation frameworks that respect cultural plurality and acknowledge the right to refuse algorithmic mediation. 
\end{abstract}



\begin{document}


%

\maketitle
\section{INTRODUCTION}
\label{introduction}
When we evaluate AI systems for fairness, there is an implicit assumption that the meaning of fairness is already established. This assumption, rarely examined yet widely adopted, forms the basis of evaluation practices that influence how AI systems are developed, deployed, and judged suitable for global use. Yet fairness is not a fixed standard but a cultural construct, interpreted differently across contexts and, at times, in ways that cannot be reconciled.

Current approaches to software fairness evaluation involve a paradox: they attempt to detect and reduce bias through frameworks that are themselves culturally situated. The metrics employed, the harms identified, and the improvements reported emerge from predominantly Western epistemological traditions. 

When assessing whether an AI system treats groups equitably, evaluators apply Western notions of equity. When measuring demographic parity, they use Western demographic categories. When examining harmful bias, they define harm through Western ethical frameworks. These evaluations therefore test whether systems align with Western ideals of fairness while presenting this alignment as universal progress.
We argue that software fairness testing, as currently practiced, operates less as a tool for justice and more as a mechanism for legitimizing culturally specific values as universal standards. 

Input to these systems comes from platforms owned and mediated by American or Western companies, leading toward bias against certain cultures {\modif{and the amplification of American cultural practices}}~\cite{li:2024:culturellm,wang:2023:not}. 
Task completion is measured against human judgments from Western 
populations \cite{atari:2023:humans}, 
and
most test datasets reflect Western contexts and concerns \cite{seth:2025:deep}. Entire knowledge systems grounded in oral traditions, Indigenous languages, and non-digital communities remain systematically excluded from both training data and evaluation frameworks.


The consequences extend beyond technical metrics. When fairness tests validate AI systems as ``unbiased'' according to narrow cultural measures, they endorse the global deployment of systems that may reproduce subtler forms of cultural harm: \emph{imposing individualistic frameworks on collectivist societies, marking non-Western problem-solving approaches as errors, or rendering Indigenous decision-making processes illegible to algorithmic systems.} These issues reflect fairness testing’s reliance on frameworks that present fairness as a universal and measurable construct, rather than recognizing it as context-dependent and culturally situated. 

At the same time, the participation required for fairness testing and the consequences of powering high-performance computing  systems raise ethical concerns for low-resource cultures \cite{li:2024:culturellm}. Furthermore, data labeling is often carried out by low-paid workers in the Global South \cite{vargas:2025:exploiting}, while the environmental costs of model training fall disproportionately on climate-vulnerable populations~\cite{regilme:2024:artificial}. 

This vision paper examines three dimensions of this problem. First, it analyses how current fairness metrics embed specific cultural assumptions while rendering others invisible (Section \ref{sub:decolonizing}). Second, it considers what datasets designed from non-Western perspectives might reveal about current blind spots (Section \ref{sub:cultural}). Third, it investigates how environmental concerns and data colonialism shape both the practice and ethics of fairness testing (Section \ref{sub:environment}). 

The paper concludes by proposing research directions that move beyond universal metrics toward evaluation frameworks that respect cultural plurality and include the possibility of refusing algorithmic mediation altogether (Section \ref{researchOpportunities}).

We include a reflection on our identities to enhance transparency \cite{desouza:2025:integrating}.
We position ourselves as researchers situated within computer science and software engineering, writing from institutions in the Global North. Our goal is not to speak for Indigenous communities but to highlight how current fairness testing practices exclude perspectives, and to argue for evaluation frameworks that allow for cultural plurality and refusal.

\section{BACKGROUND}
\label{background} 
\paragraph{Why Fairness Testing is Needed}
Software fairness testing seeks to identify that AI systems, such as large language models (LLMs), do not exhibit biases \cite{de2025software}. \textit{``Given a set of inputs, a fair software should not result in discriminatory outputs or behaviors for inputs relating to certain groups or individuals.'' }\cite[p. 1]{soremekun:2022:software}. 
Many models have been shown to exhibit biases 
related to race, gender, age, and sexual orientation \cite{simpson:2024:parity,parrish:2021:bbq,mehrabi:2021:survey}. 
This often occurs because training data contains biases, which in turn will be reflected in the predictions of the models \cite{mehrabi:2021:survey}. Systems which are built upon these models can then perpetuate inequalities and disadvantages. 
{\modif{Results can also vary depending on the language used \cite{alkhamissi:2024:investigating}, creating inconsistencies in multi-lingual contexts.}}
Because of the integration of AI systems into everyday life, people can be affected even if they are not direct users of the system (i.e., through the use of such systems for decision making) \cite{mehrabi:2021:survey,steinacker:2022:code,dehal:2024:exposing}. 
Those who finance and build the systems are more likely to belong to majoritized groups, meaning that they might lack incentives to evaluate systems for bias.
Fairness testing plays a 
role in ensuring that the software systems that affect everyday life are not completely without oversight.

\paragraph{Current Fairness Testing}
Fairness testing can focus on many different parts of the system, from the input data to compression, but most systems are not transparent to anyone outside of the company which develops it. This means that the majority of research on fairness testing focuses on opaque model testing, where inputs are given to the system and outputs are observed, without the user necessarily having knowledge of the model design or data used to train it \cite{chen:2024:fairness, de2025software}. 
{\modif{Opacity  makes it hard to observe confounding effects, for instance if competing biases are neutralizing one another.}} 

\paragraph{Concerns with Measuring Fairness}
But how do we measure fairness? \textit{Individual fairness} measures if two individuals who are similar should be classified similarly, while \textit{group fairness} considers if there is statistical parity for outcomes of groups defined by protected attributes \cite{soremekun:2022:software,chen:2024:fairness}.
However, there are a number of concerns about these approaches.
Group fairness may blatantly discriminate against individuals while maintaining statistical parity \cite{dwork:2012:fairness}. 
{\modif{Both definitions of fairness which fail to take into account the impacts of systematic bias, meaning that evaluation will prioritize equality over equity. Other concerns are feature and content production bias.}} 
Feature bias 
\modif{is} when a sensitive attribute which is not included in the data is highly correlated with another feature, and the unprotected feature is used to classify the data \cite{chen:2024:fairness}. Content production bias 
\modif{is} when prejudices based on the use of language result in variations in user experience \cite{olteanu:2019:social,mehrabi:2021:survey}.

\paragraph{Other Problems with Fairness Testing}
Quite apart from concerns related to measuring fairness, fairness testing itself is not without its critics. First, there is the question of whether trying to improve upon AI systems creates a veneer of respectability which deflects attention without providing concrete results, similar to critiques of carbon markets in relation to climate change \cite{ball:2018:carbon,netto:2020:concepts,guest:2025:against}. 
Criticisms of AI include its environmental costs, the human costs of data colonization, intellectual property theft, and the moral and intellectual effects of use \cite{guest:2025:against,regilme:2024:artificial,markelius:2024:mechanisms,arora:2023:risk,vargas:2025:exploiting,luccioni:2024:power,goetze:2024:ai,kosmyna:2025:your}.
Second, there are concerns about the effectiveness of fairness testing approaches, such as about the quality of datasets based on stereotyping \cite{blodgett:2021:stereotyping}, 
\modif{the disproportionate effects of changes in phrasing
\cite{prabhakaran:2019:perturbation,kamruzzaman:2024:woman,zhao:2024:role} which reveal potential fragility with industrial application of testing}, and the limitations of single-attribute tests to expose the interaction effects multiply-marginalized individuals experience~\cite{magee:2021:intersectional}.

A research agenda for addressing bias in AI systems called for, among other things, improving ethical design and increasing accountability, explainability and transparency \cite{desouzasantos:2025:software}. To this we add a call for a critical consideration of the \textit{fairness in fairness testing}.

\section{RETHINKING FAIRNESS TESTING}
\label{problem}

This is a vision paper focused on open issues in software fairness testing and directions for future research within software engineering. Our aim is not to prescribe definitive solutions, but to examine how fairness testing, as a branch of software testing, can evolve when embedded in diverse cultural, social, and infrastructural contexts. We argue that fairness testing must be treated as a testing practice with its own adequacy criteria, test data design, and oracles, rather than as an external evaluation added after model development. 

\subsection{Decolonizing Fairness Testing}
\label{sub:decolonizing}

A decolonized approach to fairness testing should be \textit{grounded in community input}, with the voices of underrepresented groups shaping how fairness is defined, operationalized, and tested in software systems. This means reconsidering how test cases are constructed, how adequacy is measured, and how oracles are specified in fairness evaluation. Community participation ensures that broader social perspectives are encoded into testing, 
not only in system requirements but also in the 
artifacts used to evaluate fairness \cite{begier:2010:users, soremekun:2022:software}. At the same time, fairness criteria embedded in test design may conflict because different communities introduce values that are not always aligned. For example, a fairness oracle requiring full transparency of model behavior may conflict with community knowledge that is culturally owned and not meant to be disclosed \cite{karetai:2023:decolonising, scheurich:1997:coloring}. Such conflicts should be treated as part of fairness testing itself rather than as anomalies, since they reveal how test oracles, adequacy definitions, and fairness metrics are negotiated in practice.

A decolonized view also exposes \textit{gaps in existing testing techniques}. Current fairness testing methods often rely on benchmark datasets with narrow distributions and fixed definitions of fairness, which results in underrepresented groups and localized situations being excluded from testing \cite{olteanu:2019:social, chen:2024:fairness, de2025software}. When these gaps persist, test coverage remains incomplete and reinforces systemic exclusions \cite{blodgett:2021:stereotyping, parrish:2021:bbq}. For example, speech and text models frequently misclassify dialects or non-standard language because these inputs are absent from the test data used to evaluate fairness \cite{huang:2024:understanding, seth:2025:deep}. Similarly, image datasets overrepresent some regions while leaving others invisible, meaning fairness tests cannot exercise those missing cases \cite{blodgett:2021:stereotyping}. Another issue is that fairness testing often adopts the performance of a \textit{majoritized group as the reference benchmark}, which positions deviations as deficiencies and encodes majority perspectives as the standard \cite{simpson:2024:parity, magee:2021:intersectional, nguyen2024literature}. These limitations are not just data oversights, but also testing gaps that reflect the absence of cultural and social contexts in software testing practice \cite{arora:2023:risk, regilme:2024:artificial, vargas:2025:exploiting}.

Fairness testing should 
not be reduced to a single adequacy metric or error distribution. Such reductionism implies that fairness can be fully captured by measurement, whereas testing in practice must consider \textit{subjective experiences and perceived harms}. 
This highlights how fairness failures manifest in real-world use, beyond accuracy scores. For instance, a misclassification in a dialogue system may appear trivial in test metrics but operate as a \textit{microaggression that erodes trust}; a small prediction gap in health applications may lead to \textit{serious consequences} for already disadvantaged groups \cite{sorin:2025:socio}{\modif{; and a culturally-uninformed system may provide inappropriate and risky advice~\cite{cirucci:2025:culturally}}}. Decolonizing fairness testing 
 requires broadening what counts as evidence in test oracles and recognizing fairness as a \textit{contextual adequacy goal} rather than only a technical measure.

\subsection{Embedding Cultural Perspectives in Datasets}\label{sub:cultural}

There is little doubt that embedding cultural perspectives in datasets could enhance fairness testing, in complementing the technical measures. Culture is shaped by individuals' origin, beliefs and existence over time (expressed in traditions, languages, values, and views of the world) \cite{Hofstede:2021}, and such differences are evident \modif{in} software engineering communities \cite{Zolduoarrati:2022}. 
As noted above, datasets used for model training and testing are not exempted from bias and inequality \cite{simpson:2024:parity,parrish:2021:bbq,mehrabi:2021:survey}, and there is growing awareness that the homogenising assumption that technology assumes a one size fits all, is not tenable, and is actually testing the boundaries of (un)fairness \cite{karetai:2023:decolonising}. 

However, embedding cultural perspectives in datasets presents a challenge. What would a dataset designed with multiple cultural perspectives look like? Could we localize such a dataset to specific problems/domains? 
What will form the basis of responsible data practices, accurate representations, and structural understanding of culture? Will the participation necessary from diverse communities to encode/decode \textit{their own cultural knowledge and viewpoints} be possible? What culturally situated data epistemologies and methods will facilitate the development of appropriate data representations? Who will own and govern the data?

Such datasets will need to posses specific attributes related to cultural issues, going beyond simulations and curated demos \cite{pistilli:2024:civics}, as evidence shows that efforts aimed at informing AI systems 
do not 
enhance their non-Western cultural awareness \cite{seth:2025:deep}. 

Thus, the capture of such knowledge systems (i.e., those that are not Western) could pose a challenge, as it would need participation from underrepresented communities to harvest, accurately code, validate and manage their data. Appropriate methodologies such as anthropology, sociology and philosophy of science, may help, however informed adaptations from diverse communities would be needed to align with their specific assumptions and meanings of truth. This could start with gaining the trust of these communities through genuine respectful partnerships, that centers on cultural humility and data sovereignty. 








\subsection{Implications of Environmental Concerns and Data Colonialism}\label{sub:environment}

Environmental concerns and data colonialism introduce challenges that shape both the scope and the reception of fairness testing research. One implication is that certain forms of bias may never be identified or addressed because the communities most affected by them are also those most excluded from participation in the design of testing practices. For some communities, the problem lies not only in how fairness is defined and tested, but also in whether certain AI systems should be developed or deployed at all. For example, LLMs, despite being very popular tools nowadays, are often perceived as technologies that conflict with environmental values because of their high energy consumption and resource demands \cite{luccioni:2024:power, markelius:2024:mechanisms}. In these cases, there may be little incentive to engage with fairness testing as a positive project, since the system itself is seen as misaligned with community priorities. Yet individuals may still be subject to such systems even if they personally reject them. For example, predictive tools adopted by municipalities can influence policing or surveillance practices regardless of whether local residents consent to their use \cite{regilme:2024:artificial}.  

Bias is also evident in the knowledge systems and tools that support AI. The pursuit of  `perfect' outcomes in AI systems risks overshadowing environmental awareness and reversing gains towards sustainability \cite{markelius:2024:mechanisms, ball:2018:carbon}. This is not only a technical trade-off, but also a value decision: prioritizing accuracy or performance over environmental or social concerns reflects a hierarchy of objectives. Fairness testing research must recognize that these hierarchies shape which forms of bias are prioritized for attention.  

Another implication concerns the conditions of data work. Much of the labor of data cleaning and content moderation is outsourced to workers in the Global South, who often face precarious employment and limited protections \cite{arora:2023:risk, regilme:2024:artificial}. This raises questions about whether the voices of these workers are reflected in the data. 
If their contributions are invisible or undervalued, then fairness testing 
is built on infrastructures that reproduce inequities. Data colonialism, therefore, affects both the social impact of AI development and the knowledge embedded in the systems being tested.  

Finally, global demand for AI services often relies on the intensive use of local resources, creating asymmetries where some regions bear the environmental and social costs while others benefit from the output \cite{regilme:2024:artificial,vargas:2025:exploiting}. Promoting decolonial perspectives in fairness testing, therefore, requires confronting tensions between industry priorities and community values. In particular, decolonization may conflict with measures that industry actors treat as most important, such as accuracy or efficiency. Recognizing these conflicts makes it possible to frame fairness testing not only as a technical challenge but also as a political and ethical practice that must account for environmental sustainability and global inequalities.

\section{RESEARCH OPPORTUNITIES}
\label{researchOpportunities}

We highlight the need to rethink software fairness testing by recognizing its cultural, social, and ecological dimensions. Building on the issues discussed above, we identify several directions for future research that can guide the development of more inclusive and context-sensitive evaluation practices.

\paragraph{Localized Datasets for Cultural Contexts}
Fairness testing currently relies heavily on datasets constructed from Western contexts, which means that the cultural assumptions of other regions are either invisible or misrepresented. A first opportunity lies in creating datasets localized to specific cultural and national settings{\modif{, such as KorNAT \cite{lee:2024:kornat}}}. 
Such datasets would make it possible to evaluate whether AI systems reproduce or ignore local biases and would provide a basis for identifying when fairness metrics aligned with one culture fail to capture issues in another.

\paragraph{Norm Diversity and Cultural Disagreement}
A second opportunity involves constructing evaluation datasets that explicitly capture differences in cultural norms and values. 
{\modif{The perspectives of dominant groups within the culture can be over represented by LLMs~\cite{seth:2025:deep}. A fairness testing approach that avoids a single source of truth}}
would allow researchers to examine whether AI systems present answers from a single dominant perspective or recognize situations in which multiple, sometimes conflicting, interpretations coexist. In this way, fairness testing could move from validating a single notion of correctness to assessing a model’s capacity to engage with plural perspectives.

\paragraph{Evaluating Fairness at the System Level}
Much fairness testing remains focused on isolated models, neglecting the larger systems in which these models are embedded. Future research should investigate whether fairness is being assessed only at the level of the model or across the broader sociotechnical system. Approaches such as Steinaker’s Code Capital framework \cite{steinacker:2022:code} provide one way to situate fairness within wider societal effects, including impacts on people who may be subject to systems without direct engagement. Case studies applying such frameworks could reveal forms of exclusion or harm that standard fairness metrics fail to capture.

\paragraph{Participatory and Co-Creation Approaches}
{\modif{Co-creation is already being used to create culturally-informed models such as ChatBlackGPT~\cite{egede:2025:exploring}.}}
Fairness testing is often designed and implemented without the input of those most affected by biased outcomes. There is an opportunity to integrate participatory and co-creation approaches, such as action research \cite{baum2006participatory,kemmis2008critical}, into fairness evaluation. This would mean involving underrepresented groups not merely as annotators but as collaborators in defining what fairness should mean in their contexts. Embedding community perspectives in the design of tests would make fairness evaluation more accountable and culturally grounded.

\paragraph{Adapting Models Across Cultural Contexts}
Finally, research is needed on technical methods for adapting models trained in one cultural setting to perform appropriately in another. Retrieval-augmented generation (RAG) and culturally informed prompting are possible approaches, although prompting has thus far shown limited capacity to alter biases \cite{ewart2025macro,seth:2025:deep}. The effectiveness of these approaches reducing bias requires systematic investigation \cite{li:2024:culturellm,narayan:2025:mitigating}. Techniques such as RAG, if effective, could allow fairness testing to account not only for the limitations of training data but also for the situatedness of model use across cultural contexts.

Taken together, these opportunities emphasize that fairness testing should not be reduced to universal metrics or abstract benchmarks. Instead, it should become a practice that is co-created with communities, attentive to cultural plurality, and responsive to environmental and social costs.

\section{CONCLUSION}
\label{conclusion}

This vision paper explored how software fairness testing is currently used to evaluate AI systems and identified problems in how it is framed and practiced. We showed that this practice often assumes a universal meaning of fairness that reflects Western traditions, reinforced by metrics and datasets shaped by Western institutions. These practices risk presenting context-specific values as global standards. We also discussed ethical challenges connected to the environmental costs of training and subsequent deployment of AI systems and the reliance on low-paid data labeling, which disproportionately affects vulnerable populations and concentrates risks in the Global South.  

We identified three domains where these problems can be addressed. Decolonizing fairness testing can support a shift away from universal definitions toward practices shaped by different communities, treating fairness as a localized and negotiated property rather than a fixed measure. Embedding cultural perspectives in dataset development can help reduce exclusions and blind spots created by dominant benchmarks, allowing fairness testing to better reflect linguistic, social, and cultural diversity. Addressing environmental concerns and data colonialism requires recognizing that fairness testing is embedded in infrastructures that carry ecological and social costs, which are unevenly distributed across global contexts.  

By reflecting on these dimensions, we conclude that fairness testing is not only a technical challenge but also a political and ethical one. Future research must investigate how decolonial approaches can be operationalized in testing practices, how datasets can be designed to represent plural cultural perspectives, and how fairness research can incorporate sustainability and labor considerations without being subordinated to efficiency or accuracy alone. As a vision paper, our contribution lies in identifying these directions as open opportunities for the software engineering and AI communities, emphasizing that fairness testing should evolve toward an inclusive and contextual practice that accounts for cultural, social, and ecological realities in AI development.

\bibliographystyle{acm}
\bibliography{paper}

\end{document}